\documentclass[aps,pra,superscriptaddress,twocolumn]{revtex4-2}
\usepackage{graphicx}
\usepackage{dcolumn}
\usepackage{bm}
\usepackage{nicefrac}
\usepackage{amsfonts,amsmath,amssymb,stmaryrd}
\usepackage{braket}
\usepackage{tabularx}
\usepackage{multirow} 
\usepackage{hhline}
\usepackage{subfigure}  
\usepackage{bbm} 
\usepackage[pdftex]{epsfig}
\usepackage{mathrsfs}
\usepackage{verbatim}
\usepackage{ulem}
\usepackage{array}
\usepackage{cancel}
\usepackage{ifthen}
\usepackage{float} 
\usepackage{listings}
\usepackage{color}
\usepackage{hyperref}
\hypersetup{colorlinks=true,linkcolor=blue,anchorcolor=blue,citecolor=blue,filecolor=blue,urlcolor=blue}
\let\revappendix\appendix
\usepackage[nameinlink,noabbrev]{cleveref}

\begin{document}

\title{Quantum description of atomic diffraction by material nanostructures}

\author{Charles Garcion}
\affiliation{Laboratoire de Physique des Lasers, Université Sorbonne Paris Nord, CNRS UMR 7538, F-93430, Villetaneuse, France.}
\affiliation{Leibniz University of Hanover, Institute of Quantum Optics, QUEST-Leibniz
Research School, Hanover, Germany.}

\author{Quentin Bouton}
\affiliation{Laboratoire de Physique des Lasers, Université Sorbonne Paris Nord, CNRS UMR 7538, F-93430, Villetaneuse, France.}

\author{Julien Lecoffre}
\affiliation{Laboratoire de Physique des Lasers, Université Sorbonne Paris Nord, CNRS UMR 7538, F-93430, Villetaneuse, France.}

\author{Nathalie Fabre}
\affiliation{Laboratoire de Physique des Lasers, Université Sorbonne Paris Nord, CNRS UMR 7538, F-93430, Villetaneuse, France.}

\author{Éric Charron}
\affiliation{Université Paris-Saclay, CNRS, Institut des Sciences Moléculaires d’Orsay, F-91405 Orsay, France.}

\author{Gabriel Dutier}
\affiliation{Laboratoire de Physique des Lasers, Université Sorbonne Paris Nord, CNRS UMR 7538, F-93430, Villetaneuse, France.}

\author{Naceur Gaaloul}
\affiliation{Leibniz University of Hanover, Institute of Quantum Optics, QUEST-Leibniz Research School, Hanover, Germany.}

\date{\today}

\begin{abstract}
We present a theoretical model of matter-wave diffraction through a material nanostructure. This model is based on the numerical solution of the time-dependent Schrödinger equation, which goes beyond the standard semi-classical approach. In particular, we consider the dispersion force interaction between the atoms and the material, which is responsible for high energy variations. The effect of such forces on the quantum model is investigated, along with a comparison with the semi-classical model. In particular, for atoms at low velocity and close to the material surface, the semi-classical approach fails, while the quantum model accurately describes the expected diffraction pattern. This description is thus relevant for slow and cold atom experiments where increased precision is required, e.g. for metrological applications.
\end{abstract}
\maketitle



\section{\label{sec:level1}Introduction}

Atomic interferometry using light pulses \cite{Geiger2020}, magnetic gradients \cite{Machluf2013}, and material gratings \cite{Cronin2009} is now a mature field of physics. In particular, advances in the cooling and control of atoms have turned this field into a versatile tool for precise measurements with applications in fundamental physics tests \cite{Tino2021} or accurate inertial sensing \cite{Dutta2016}. In this work, we focus on atomic diffraction patterns formed by material gratings where dispersion forces such as Casimir-Polder play an important role.   

These forces between the particle and the grating walls are created by the ground state fluctuations of the electromagnetic fields, generally resulting in an attractive potential \cite{Buhmann2012}. They are of far-reaching importance in chemistry, biology, cosmology, atomic force microscopy \cite{Bruch1997} and can be used as a test of quantum electrodynamics. Interest in understanding this interaction also stems from its application, e.g in the development of atomic lithography \cite{Fiedler2022}. The role of this interaction is all the more important since the mechanical gratings used in atomic interferometry experiments are usually built on the nanometer scale. Using nano-fabricated transmission gratings (called nanograting), the influence of this force has been studied for alkali atoms \cite{Perreault2005,Lepoutre2009}, excited noble gases \cite{Grisenti1999, Bruhl2002} or even complex molecules \cite{Brand2015}. Furthermore, this approach has recently allowed to distinguish between the non-retarded and the retarded regime of the Casimir-Polder interaction in the intermediate range \cite{Garcion2021}. Such interferometers can also be configured  to measure, for example, atomic polarizability with high sensitivity \cite{Gregoire2015}, dynamic polarizability of large molecules \cite{Hackermuller2007} and inertial signals \cite{Trubko2015}.

These matter-wave experiments require quantitative simulations to bridge the gap between theory and experiment. The current theoretical framework used so far to account for the dispersion forces is based on a semi-classical approach in the eikonal approximation \cite{Cronin2004} and beyond \cite{Nimmrichter2008}. This method remains valid if the action $S$ over the classical trajectories through the interaction region is much larger than $\hbar$, if the de Broglie wavelength $\mathrm{\lambda}$ of the particle is much shorter than the spatial variation of the interaction potential $V$, and if the spatial variation of the de Broglie wavelength is small. These approximations fail in close vicinity of the walls, where the dispersion force is dominant, and for low atom velocities. In both cases, the interaction potential exceeds the kinetic energy. Triggered by $(i)$ the increased precision required in metrology, $(ii)$ the need to probe the dispersion forces close to the surface, and $(iii)$ the experimental progress in cooling and slowing atoms to increase the interaction time between the atoms and the nanograting thus enhancing the sensitivity of the measurements \cite{Garcion2021}, it becomes necessary to advance the theoretical description of material matter-wave interference beyond the semi-classical approach.  

We present here a matter-wave diffraction model based on a numerically efficient solution of the time-dependent Schrödinger equation. Owing to the fast developments in the field of matter-wave optics in the last decade, numerical simulations involving atom-surface interactions have been developed in the context of quantum reflection \cite{Herwerth2013,Galiffi2017}. Here, by exploring short atom-surface distances where large energy variations occur, we reproduce matter-wave diffraction patterns beyond the semi-classical approach.  

The structure of the paper is as follows: in Section~\ref{sec:level2}, we describe the theoretical approach, introducing the dispersion forces used and the approximation adopted in the immediate vicinity of the surface. Section~\ref{sec:level3} is devoted to the numerical performance and to the description of the diffraction pattern in the far-field regime. In Section~\ref{sec:level4}, the numerical results obtained are compared with the semi-classical approach. Finally, a summary and conclusion are given in Section~\ref{sec:level5}.

\section{\label{sec:level2}Theoretical model}
\subsection{Representation of the problem}

\begin{figure*}
    \centering
    \includegraphics[width=15cm]{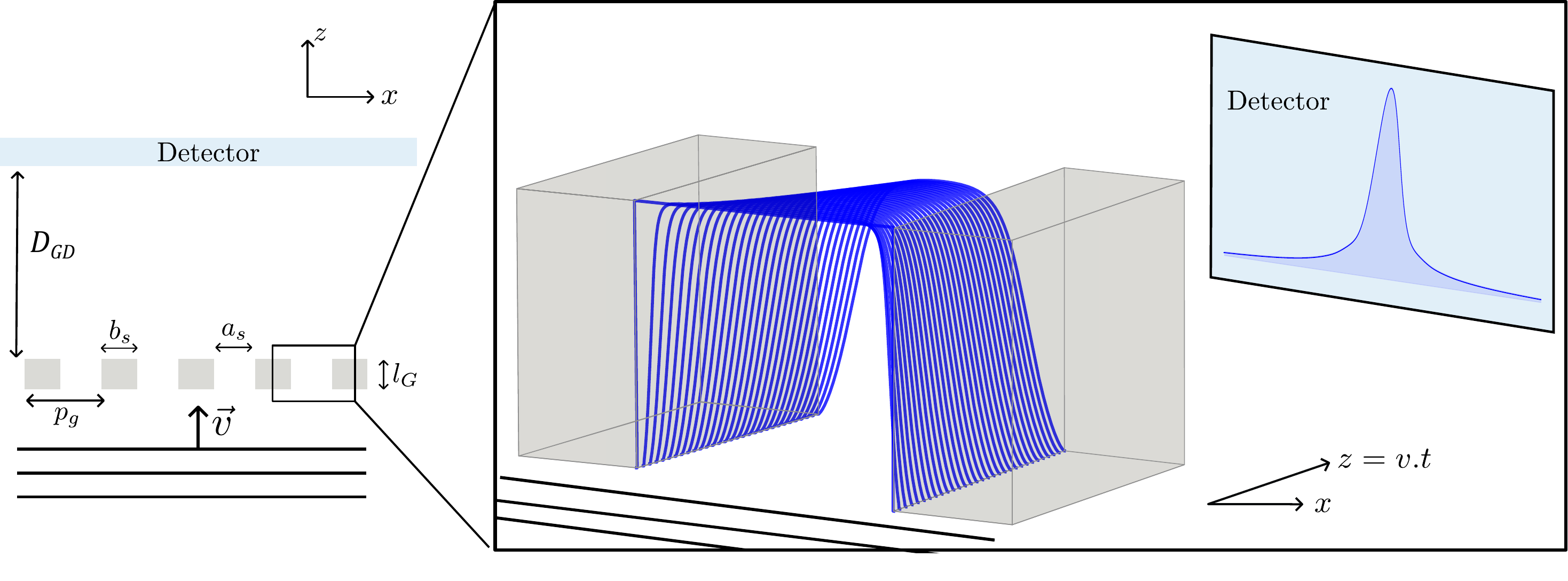}
    \caption{On the left, a schematic illustration of the problem considered: a plane wave representing a single particle with velocity $v$ hits a $N$-slit nanograting. The detection takes place at a distance $D_{GD}$ from the slits. The nanograting has the following geometry: slit size $a_s$, wall slit $b_s$, thickness $l_G$ and periodicity $p_{g}=a_s+b_s$. On the right is the result of a typical time evolution of the wave function inside the slits, with $a_s$=100 nm, $b_s$ = 100 nm and $l_G$ = 100 nm, which are typical dimensions used in experiments. In the back plane a characteristic one-slit wave function is plotted after free propagation to the detector. This function is the envelope of the $N$-slit wave function.}
    \label{fig:Grating-1slit V2}
\end{figure*}

In this section we outline the matter-wave interference pattern from a $N$-slits nanograting in the far-field limit and show that the problem can be reduced to a single-slit problem. The system studied is illustrated schematically in Fig.~\ref{fig:Grating-1slit V2}. The coordinate system is chosen such that the grating lies in the $(xy)$ plane and $z$ is the propagation direction. The incoming atoms, of mass $m$, are described by an incident plane wave $\psi_{\mathrm{inc}}(z) = \exp(ikz)$ with $k = 2 \pi/ \lambda$, where $\lambda$ is the de Broglie wavelength of the atoms. This incident plane wave interacts with a $N$-slits nanograting of slit size $a_s$, wall width $b_s$ and thickness $l_G$ (see Fig.~\ref{fig:Grating-1slit V2}). We also assume that the slits are large enough along the $z$-axis to ignore the diffraction pattern along this given direction. The grating periodicity $p_g= a_s+b_s$ implies that the total wave function inside the grating can be expressed as the given quantum superposition  
\begin{equation}\label{eqn:Psi_propag_inside_slits}
    \psi(x,t) = \sum_{n=0}^{N-1} \psi_{n}(x,t) ,
\end{equation}
where $\psi_{n}(x,t)$ represents the wave function of the particle inside the $n$-th slit. We also assume that the slits are independent and identical, allowing us to write $\psi_{n}(x,t) =  \psi_{0}(x-n p_{g},t)$. After an interaction time $t_{e}=l_{G}/ v $ inside the grating, where $v=\hbar k/m$ is the particle velocity, the Fourier transform of the total wave function at the exit of the grating is simply given by
\begin{equation}
\label{eqn:Psi_propag_inside_slits_TF}
\Tilde{\psi}(k,t_e) = \Tilde{\psi_{0}}(k,t_e) \left[\frac{\sin\left(\frac{N k p_{g}}{2}\right)}{\sin\left(\frac{k p_{g}}{2}\right)}\right] e^{i\frac{(N-1)}{2}k p_{g}}.
\end{equation}
Thereafter, the atoms freely evolve to the detector during a time $T$, leading to the following expression of the propagated wave function in the Fourier domain   
\begin{equation}
\Tilde{\psi}(k,t_e+T) = \Tilde{\psi}(k,t_e)\;e^{-i\,\frac{\hbar k^2}{2m}T}.
\end{equation}
For detection in the far-field regime, i.e. after a long propagation time, we use the stationary phase approximation (see Appendix~\ref{app:level1} and e.g. Ref.~\cite{Mandel1995}) to calculate the wave function in the coordinate space, which can be expressed at the detector as 
 \begin{equation}
  \psi(x,t_e+T) \simeq \sqrt{\frac{m}{\hbar T}}\;\,\Tilde{\psi}\!\left(\frac{m x}{\hbar T},t_e \right)\,e^{i\frac{m x^2}{2\hbar T}}\;e^{-i\frac{\pi}{4}}.
\end{equation}
In the typical cases that we are going to encounter in this study, this approximation yields negligible error, the error bound with the second order correction being below $2 \times 10^{-4}$ (see Appendix~\ref{app:level1}). It also avoids scaling problems in terms of number of grid points required to numerically compute the far-field diffraction pattern. Finally, under this approximation the final probability density can be simply expressed as 
\begin{equation}\label{eqn:Psi_Nslit_Theta0}
|\psi(x,t_{e}+T)|^2 \simeq  \frac{m}{\hbar T}\;\left[\frac{\sin\left(N k_g\,x\right)}{\sin\left(k_g\,x\right)}\right]^2 |\psi_{0}(x,t_{e})|^2
\end{equation}
where
\begin{equation}
k_g = \frac{m p_g}{2\hbar T}\,.
\end{equation}
In this expression, $|\psi_{0}(x,t_{e})|^2  $ is the atomic probability density diffracted by a single slit. In the limit of large slit numbers ($N \gg 1$), this expression can be further simplified to
\begin{equation}\label{eqn:Psi_Nslit_Theta0_largeN}
     |\psi(x,t_{e}+T)|^2 \propto |\psi_{0}(x,t_{e})|^2  \sum_{i} \delta\!\left(x-i\frac{\lambda D_{GD}}{p_{g}} \right)  
\end{equation}
where $\delta(x)$ is the Dirac delta function and $D_{GD}$ the grating-detector distance. As a result, the problem in the far-field regime, where the wave pattern is expanded well beyond the size of the slit $a_{s}$, is reduced to simulating the propagation through a single slit.

\subsection{Potential inside the slits}

The existence of forces between the atoms and the slit surface is due to the Casimir-Polder interaction \cite{Sipe1984,Sipe1985}. In this article, we consider the dispersion force inside the slits for atoms in the vicinity of a surface, neglecting the retardation effects. In this case, the potential, also called the non-retarded Casimir-Polder or van der Waals interaction, is written for the two walls of the slits located at $x= \pm a_{s}/2$ as
\begin{equation}\label{eqn:Potential_CP}
V_{vdW}(x) = -\frac{C_3}{\left( a_{s}/2-x\right)^3} -\frac{C_3}{\left( a_{s}/2+x\right)^3},
\end{equation}
where $C_{3}$ is a coefficient describing the strength of the interaction. This coefficient depends on the polarizability of the atom and on the dielectric response of the wall material. Inspired by the experimental setup reported in \cite{Garcion2021}, we consider Argon atoms in the $^{3}P_{2}$ state interacting with a $\mathrm{Si_{3}N_{4}}$ grating, giving $C_{3}=5.04$ meV.nm$^3$ \cite{C3_calcul}. When the wave functions of the atoms start to overlap with the electron of the surface, an additional short-range repulsive contribution arises due to the Pauli repulsion. We model such a contribution with the repulsive part of a 9-3 Lennard-Jones potential for each wall of the slit \cite{Bryk1999, Maury2016}
\begin{equation}\label{eqn:TDSE_V_AS_form}
    V_{LJ}(x)=\frac{C_{rep}}{\left( a_{s}/2-x\right)^9} +\frac{C_{rep}}{\left( a_{s}/2+x\right)^9},
\end{equation}
where $C_{rep}$ is a strength coefficient. The total potential is thus $V_{TOT}(x)= V_{vdW}(x) + V_{LJ}(x)$ and is represented in Fig. \ref{fig:Pot_and_mask_function}(a). In the following, we fix the position of the minimum potential at a distance $r_{min}$ from the walls. This implies that the coefficients $C_{3}$, $r_{min}$ and $C_{rep}$ are no longer independent and are related by $3\,C_{rep} = C_{3}\,r_{min}^{6}$. For the simulation, we use $r_{min}$ = 0.35 nm, which corresponds to the radius of $^{3}P_{2}$ Argon atoms in a solid sphere model (see Appendix~\ref{app:level2}).

\subsection{Adjustment of the potential and wave function absorption}

\begin{figure}[t]
{\includegraphics[width=\linewidth]{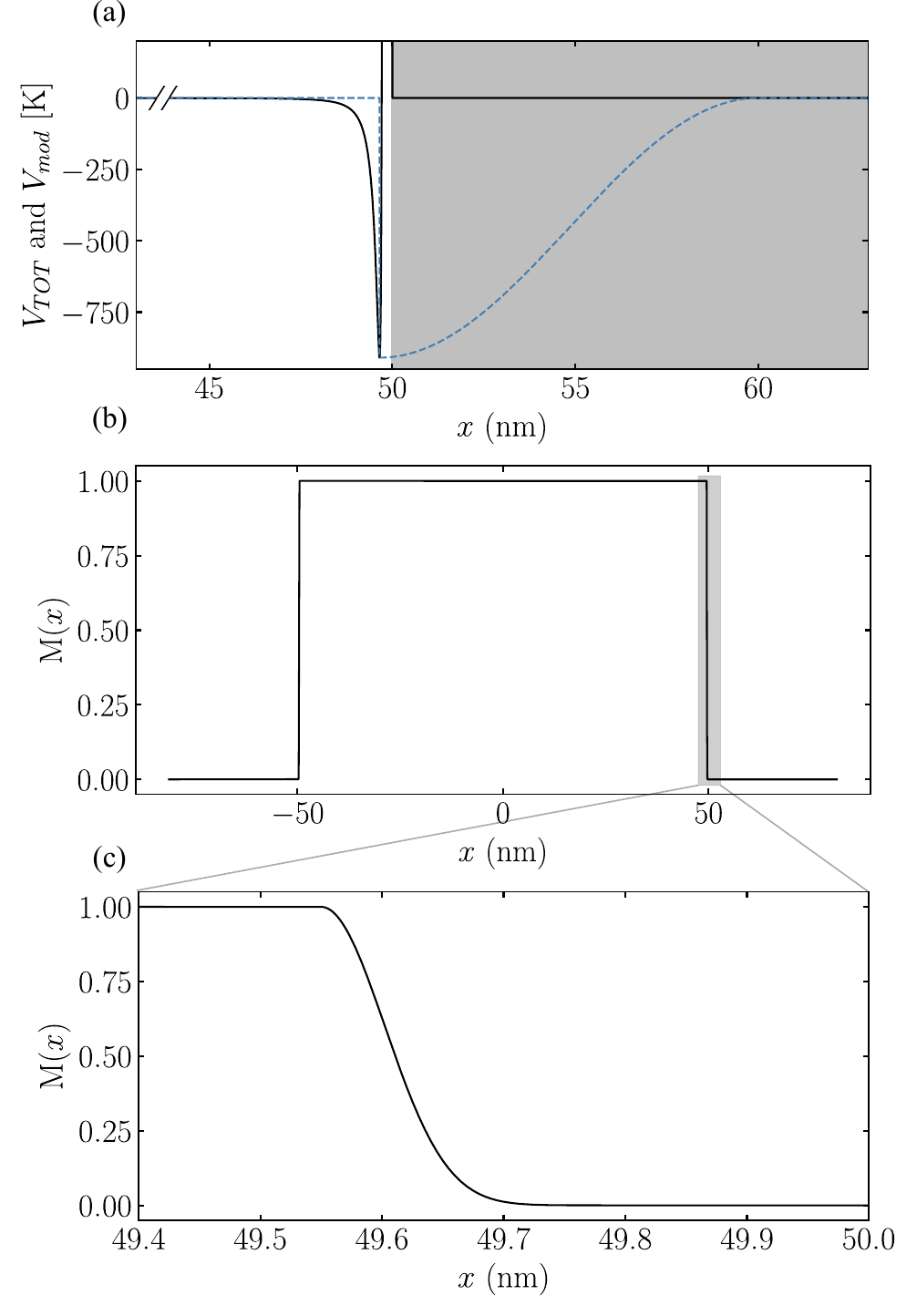}}
 \caption{\label{fig:Pot_and_mask_function}%
 (a) Total potential $V_{TOT}(x) = V_{vdW}(x) + V_{LJ}(x)$ (solid black line) as a function of the distance $x$ inside the slits. The grey area represents a wall located at $x = a_{s}/2$ = 50 nm. The blue dashed curve represents the modified potential $V_{mod}(x)$. (b) Mask function $M(x)$ used to absorb the wave function close to the surfaces. Here $d=0.2 \ \text{nm}$ and $x_{abs}=49.75\ \text{nm}$ (c) Zoom on the mask function $M(x)$.}%
\end{figure}

As seen in Fig.~\ref{fig:Pot_and_mask_function}(a), the overlapping between the wave functions of the atoms and the surface occurs for atom-surface distances shorter than $r_{min}$, leading to sticking processes \cite{Lonij2009} or internal state transfer \cite{Vassen2012} for the atoms. In all cases, such atoms are lost and do not contribute to the diffraction pattern signal. Wave function reflections for the short-range regime are thus annihilated, and an absorption method must be implemented. Hence, an adjustment of the potential $V_{TOT}$ is introduced in the model \cite{Herwerth2013}. The underlying idea is to replace the repulsive part of the potential to improve the absorption effect. In the simulation we use the modified potential $V_{mod}(x)$ defined as follows. Inside the slit and far from the surfaces, i.e. for $|x| \leqslant a_{s}/2-r_{min}$, $V_{mod}(x) = V_{TOT}(x)$. In the vicinity of the slit surface, i.e. for $a_{s}/2-r_{min} \leqslant |x| \leqslant a_{s}/2+l_{ab}$
\begin{equation}
V_{mod}(x) = U_0\,\cos^2\!\left[\frac{\pi\big(|x|-a_{s}/2+r_{min}\big)}{2\big(l_{ab}+r_{min}\big)} \right]
\end{equation}
where $U_0 =  V_{TOT}(a_{s}/2-r_{min})$ is the minimum of the potential and $l_{ab}$ is the distance over which the potential $V_{mod}(x)$ goes from $U_{0}$ to 0 (see Fig.~\ref{fig:Pot_and_mask_function}). And finally, for $|x| \geqslant a_{s}/2+l_{ab}$, $V_{mod}(x) = 0$. A typical value for $l_{ab}$ in the simulation is $l_{ab}$ = 10 nm. This modified potential, as well as its derivatives, are continuous at the positions $|x| = a_{s}/2 - r_{min}$. 

To account for atomic losses near the boundaries $|x| = a_{s/2}-r_{min}$, we introduce a mask function $M(x)$ to impose absorbing boundary conditions. In practice, this is done in a simple way since the wave function $\psi(x,t)$ is multiplied by $M(x)$ after each time step of the propagation. This absorption technique is similar to other absorptions methods such as the imaginary negative potential \cite{Kosloff1986} or the complex absorption potential \cite{Hussain2000}. The mask function has been optimized for our specific case. It is defined as follows. Inside the slit and far from the surfaces, i.e. for $|x| \leqslant x_{abs}-d$, $M(x) = 1$. Here $x_{abs}$ is the position where the wave function is absorbed and $d$ is half the total absorption length. In the absorption region, i.e. for $x_{abs}-d \leqslant |x| \leqslant x_{abs}+d$ we have
\begin{equation}
M(x) = \cos^{12}\!\left[\frac{\pi\big(|x|-x_{abs}+d\big)}{4d}\right].
\end{equation}
And for $|x| \geqslant x_{abs}+d$, $M(x)=0$. The parameter $x_{abs}$ is chosen so that the absorption occurs at a correct distance from the wall. For the simulations we use $d=0.2$ nm and $x_{abs}=49.75$ nm (see Fig.~\ref{fig:Pot_and_mask_function}(b) and (c)). In addition, we choose the initial wave function at the entrance of the slit to be proportional to the mask function.

\section{\label{sec:level3}Numerical method}

The 1D time-dependent Schrödinger equation is solved numerically using the second-order split-operator technique \cite{Feit1982}. For the simulations performed in this section, the initial atom velocity is set to $v=$ 20 m/s and the nanograting geometry properties are $a_{s} = $ 100 nm, $b_{s}$ = 100 nm and $l_{G}= $ 100 nm. The spatial grid resolution is $\delta x=$ 2.5 pm and the time step is $\delta t=$ 0.25 ps, values for which the simulations numerically converged. The number of grid points used is $2^{16}$. The grating-detector distance is fixed at $D_{GD}=$ 307 mm such as the detection takes place in the far-field regime with a Fresnel number $\mathcal{F}= a_{s}^2 / (\lambda D_{GD}) \ll 10^{-3}$. 

\subsection{Probability density inside the slit}

\begin{figure}[t]
 \includegraphics[width=\linewidth]{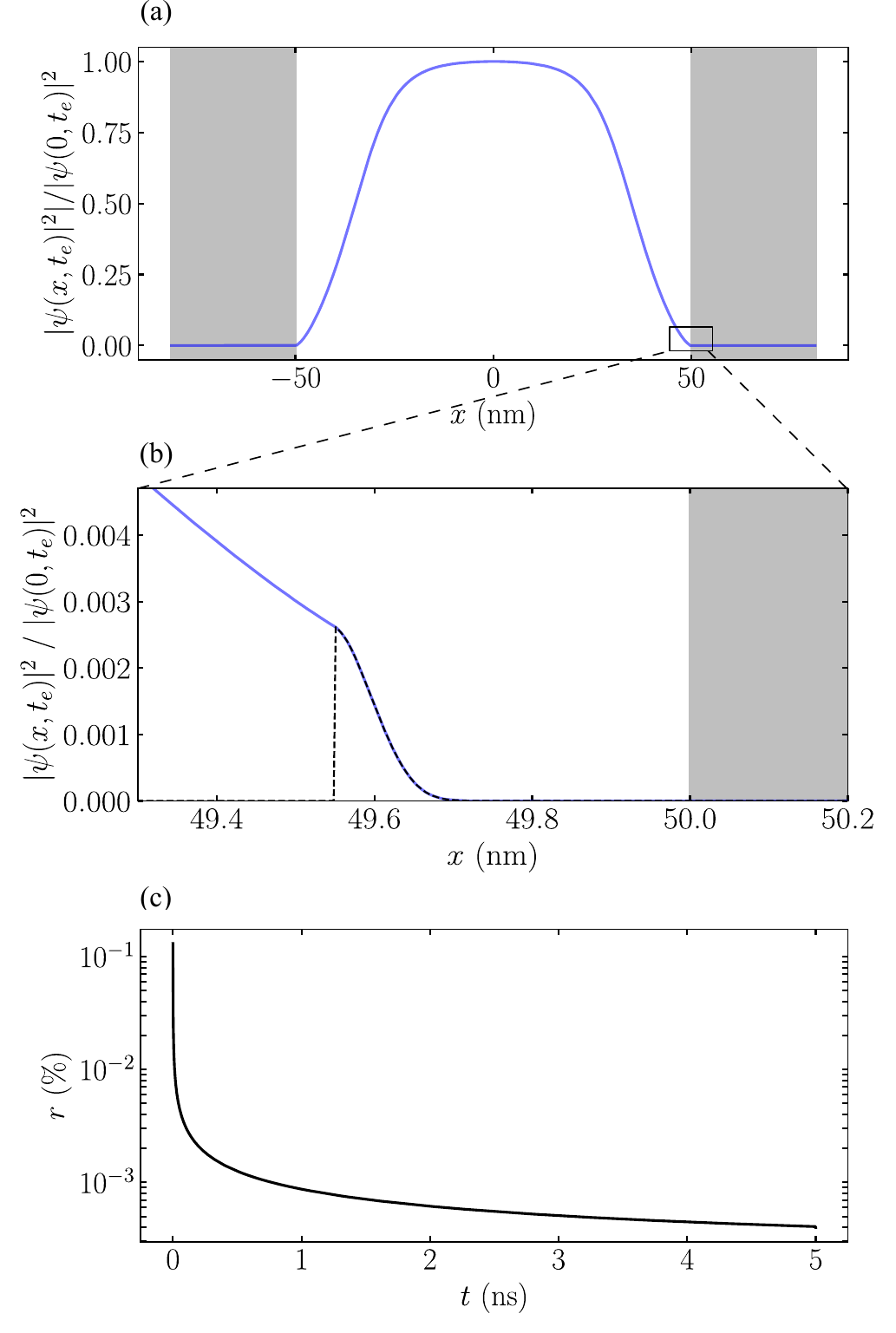}
 \caption{\label{fig:influence_of_mask}%
 (a) In the solid blue line, the square modulus of the wave function $|\psi(x,t_{e})|^{2}$ at the exit of the slit. The two grey areas represent the slit walls. In (b), zoom on $|\psi(x,te)|^{2}$ close to the surface. The black dashed line indicates the modification area of the wave function due to the absorption region ($x \geqslant x_{abs}-d$). (c) Time evolution of the ratio $r(t)$ in a log-scale.}%
\end{figure}

Figure \ref{fig:influence_of_mask} shows the square modulus of the wave function at the exit of the slit $|\psi_{0}(x,t_{e})|^{2}$. We observe that the probability of finding an atom near the surface is low. This is due to atom losses when hitting the surface during the propagation. We can also observe that the absorption of the wave function occurs over a length of about $0.1\ \text{nm}$, which is smaller than the absorption length $d=$ 0.2 nm. We can also verify that the damped part of the wave function is centered around $49.65\ \text{nm}$, which corresponds to $r_{min}=0.35\ \text{nm}$.

The mask function and the associated absorbing length $d$ affect the shape of the wave function $\psi_{0}(x,t)$ near the surface where the absorption occurs. Therefore, to quantify the influence of both the absorbing length $d$ and the modified potential $V_{mod}(x)$ on the wave function $\psi(x,t)$, we introduce the ratio $r(t)$ between the integral of the square modulus of the wave function in the absorbing region, with the integral of the square modulus of the total wave function 
\begin{equation}\label{eqn:ratio}
r(t)=\frac{\displaystyle 2\int_{x_{abs}-d}^{+\infty}|\psi_{0}(x,t)|^2 \mathrm{d}x}{\displaystyle\int_{-\infty}^{+\infty}|\psi_{0}(x,t)|^2 \mathrm{d}x}.
\end{equation}
At the entrance of the slit (time $t=0$) we have $r(0) \simeq 0.125  \ \%$ and at the exit ($t=t_e$) we have $r(t_{e}) \simeq 4.1\times10^{-4}\ \%$. The decrease of $r(t)$ is explained by the reduction of the amplitude of the wave function near the walls due to atoms losses. The small initial value and rapid decrease of $r(t)$ (see Fig.~\ref{fig:influence_of_mask}(c)) justifies that the absorbing length $d$ is sufficiently small to have a negligible influence on the wave function at the exit of the slit. Furthermore, it also supports the statement that the modification of the repulsive part of the potential has a minor impact.

\subsection{Probability density at the detector and incoherence effects}

\begin{figure}[!t]
    \centering
    \includegraphics[width=\linewidth]{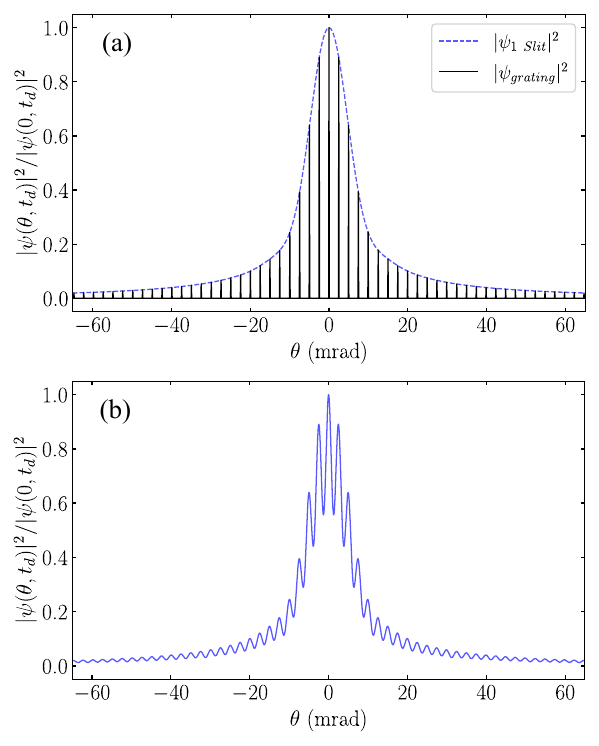}
    \caption{(a) Square modulus of the wave function $|\psi(x,t_{e}+T)|^{2}$ at the detector for the diffraction by one slit $|\psi_{1 Slit}|^{2}$ (dashed blue line) and by $N \gg 1$ slits $|\psi_{grating}|^{2}$ (solid black line) as a function of the diffraction angle $\theta= \arctan(x/D_{GD}) \simeq x/D_{GD}$. Both functions are related to Eq.~(\ref{eqn:Psi_Nslit_Theta0_largeN}), where $|\psi_{1 Slit}|^{2}$ is the envelope for $|\psi_{grating}|^{2}$, in a similar way as in wave optics. (b) Example of the expected diffraction pattern of a $N$-slits nanograting obtained with our numerical simulations taking into account incoherence effects. The chosen angular beam divergence is $\sigma = 0.8$ mrad.}
    \label{fig:psi_1slit_and_Nslits}
\end{figure}

After an interaction time $t_{e}$ with the grating, the wave function propagates freely to the detector for a time $T$. The result is shown in Fig.~\ref{fig:psi_1slit_and_Nslits} for the diffraction by one slit and by $N$-slits ($N \gg 1$).  

\begin{figure*}
    \centering
    \includegraphics[width=\linewidth]{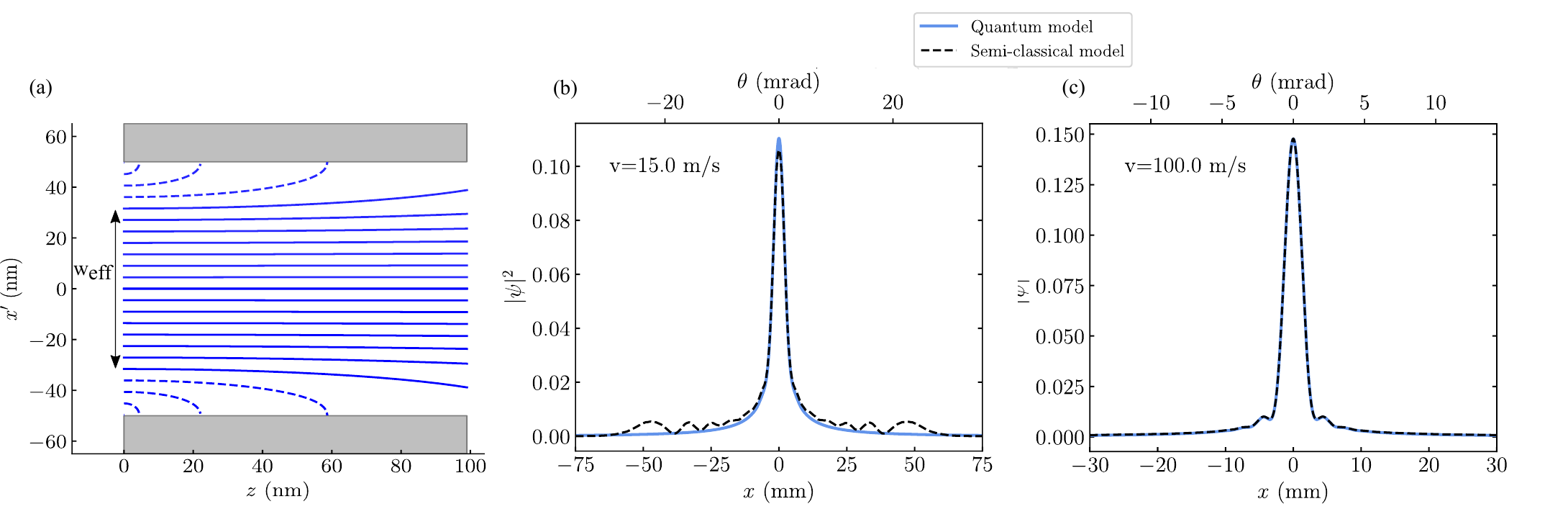}
    \caption{Comparison between the single-slit diffraction pattern derived from the numerical simulation and the semi-classical model. The numerical simulations are performed with grating wall size $a_{s}$=100 nm, grating thickness $l_{G}$=100 nm, $C_{3}=5.04$ $\mathrm{meV.nm^{3}}$ and a propagation time from the grating to the detector $T=$ 21 ms. These parameters ensure that the far-field regime is reached. In (a), an example of classical particles motion governed by Newton's law $m \ddot{r} = - \nabla V_{vdW}$. The velocity of the incoming particles is set to $v= 15$ m/s. To include the absorption effect in the simulation, atoms reaching the coordinate $x^{\prime} = \pm a_{s}/2 - r_{min}$ during their propagations inside the slit are removed (dashed lines). This corresponds to an effective slit width $\mathrm{W_{eff}}$ \cite{Grisenti1999}. The fraction $F$ of atoms leaving the slits is thus given by $F=\mathrm{W_{eff}/a_{s}}$. In (b) and (c) we compare the normalized diffraction pattern obtained by the semi-classical approach with the numerical simulation for atomic velocities of $v=$ 15 m/s and  $v=$100 m/s. We find a good agreement for the high velocity $v=100$ m/s. However, at the low velocity $v$=15 m/s a significant disagreement is observed.}
    \label{fig:trajectories_semi_classic}
\end{figure*}

In order to simulate a real experiment, incoherence effects must be taken into account. In general, the incoherence effects come from the spatial extension of the source, related to the van Cittert–Zernike theorem, and from the non-monochromaticity of the atomic source, related to the Wiener–Khinchin theorem, leading to longitudinal $k_{z}$ and transversal $k_{x}$ momenta distributions. We can describe the spatial incoherence effects by convolving the ideal case of the diffraction by $N$-slits $|\psi(\theta \simeq x / D_{GD},t_{e}+T)|^{2}$ by a Gaussian distribution  
\begin{equation}\label{eqn:gauss}
 G(\theta)=\frac{1}{\sqrt{2\pi\sigma^2}} \exp{ \left( {-\frac{\theta^2}{2\sigma^2}} \right) },
\end{equation}
where $\theta = \arctan(x/D_{GD})$ is the diffraction angle and $\sigma$ the standard deviation of the distribution. In a typical experiment, $\sigma$ does not exceed 1 mrad, corresponding to a well-collimated beam. The final diffraction pattern is thus $I(\theta,t_{e}+T) = |\psi (\theta,t_{e}+T)|^{2} * G(\theta)$ and is shown in Fig.~\ref{fig:psi_1slit_and_Nslits} for an angular beam distribution of $0.8\ \text{mrad}$. The angular beam distribution being small, its influence on the wave packet propagation at normal incidence is negligible.

\section{\label{sec:level4}Comparison with the semi-classical approach}

In this section, the numerical results obtained are compared with the commonly used semi-classical approach. For the comparison, we will only consider the diffraction by one slit (which is related to the $N$-slits diffraction by Eq.~(\ref{eqn:Psi_Nslit_Theta0})) and we will not include the incoherence effects. The geometrical parameters of the nanogratings remain the same as in the previous sections, as well as the strength of the Casimir-Polder interaction parameter $C_{3}$. In the semi-classical approach, the diffraction amplitude at the detector $\psi_{SC}(x,t_{e}+T)$ is described with classical waves, which can be developed trough Kirchhoff's diffraction formula \cite{Born1999} 
\begin{align}\label{eqn:Kirchhoff}
\psi_{SC}(x,t_{e}+T)& \propto \int  \left(\frac{\cos\mathcal{V}+\cos\mathcal{V}'}{2\lambda}\right) \psi_{0}(x',t_{e}) \nonumber \\
& \qquad\qquad \times \; \exp\left( i \frac{2\pi x x'}{\lambda D_{GD}} \right) dx'\,,
\end{align}
where $\psi_{0}(x^{\prime},t_{e})$ is atomic wave function at the output of the slit, $ \mathcal{V} =\arctan(|x-x'|/D_{GD})$ and $\mathcal{V}' = \arctan(v_{x}'/v_{z})$ are the geometric correction angles. This expression is valid in the Fresnel approximation, i.e. as long as the propagation distance $D_{GD}$ satisfies the inequality $D_{GD} \gg a_{s}$. This condition is met in the simulations since we study the diffraction pattern in the far-field regime with a Fresnel number $\mathcal{F} \ll 10^{-3}$. In the framework of the semi-classical treatment of the atomic center of mass motion, referred as the time dependent quasi-classical approximation, the atomic wave function is estimated by means of action integrals along classical trajectories \cite{Henkel1994}. In this approximation, the phase varies very rapidly along the different possible paths of the interferometer, and most of the interference will be destructive, except for the classical path. In this case, the wave function $\psi_{0}(x^{\prime},t_{e})$ writes
\begin{equation}
\psi_{0}(x^{\prime},t_{e}) = \exp \left( \frac{i}{\hbar} S(x^{\prime},t_{e}) \right),
\end{equation}
where
\begin{equation}\label{eqn:action}
S(x^{\prime},t_{e}) = \int_{0}^{t_{e}} L(t) \mathrm{d}t
\end{equation}
is the action integral along the classical trajectories. Developing Eq.~(\ref{eqn:action}) we have 
\begin{align}\label{eqn:action2}
\psi_{0}(x^{\prime},t_{e})&=\exp \left( \frac{i}{2 \hbar} m v_{z}^{2} t_{e} \right) \times  \exp \left( \frac{i m}{2 \hbar} \int_{0}^{t_{e}} v_{x^{\prime}}^{2}(t) \mathrm{d}t \right)  \nonumber \\
& \qquad\qquad \times \exp \left( -\frac{i}{ \hbar} V_{vdW}(x^{\prime})\, t_{e} \right)
\end{align}
where $v_{z}$ (respectively $v_{x^{\prime}}$) is the particle velocity in the $z$ (respectively $x^{\prime}$) axis inside the slits. The first term, which is independent of $x^{\prime}$, acts as a global phase and thus does not have a contribution to the diffraction pattern. The second term allows to go beyond the eikonal approximation, i.e. when the atomic trajectories are not constrained on the $z$ axis. In fast atom beam experiments, with velocities between 200 m/s and 2000 m/s, this term is generally almost constant and is thus neglected in the phase estimation \cite{Perreault2005,Lepoutre2009,Grisenti1999}. 

\begin{figure}[!t]
    \centering
    \includegraphics[width=\linewidth]{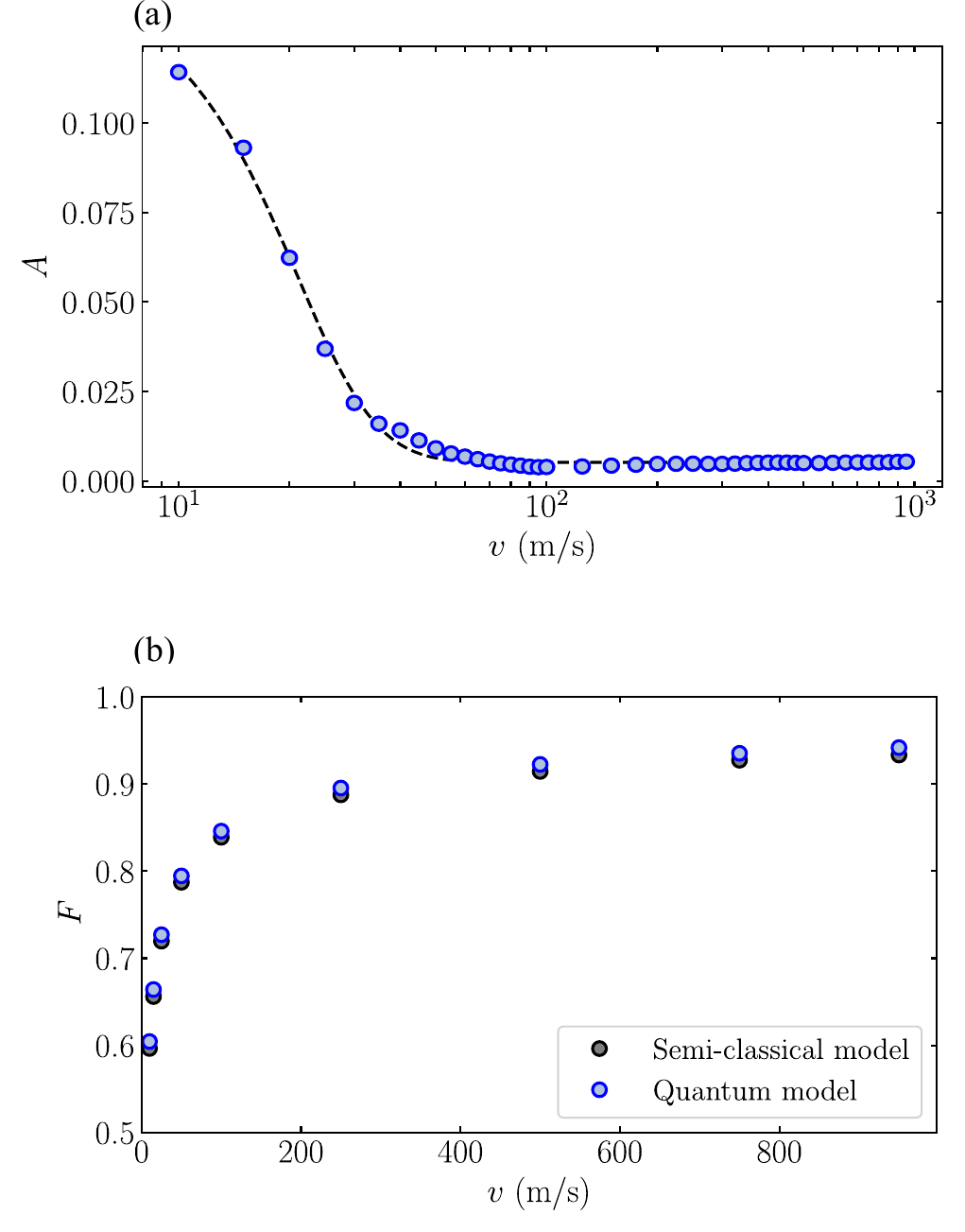}
    \caption{(a) Dimensionless quantity $A$ as a function of the velocity $v$ of the incoming atoms. $A$ quantifies the relative difference between the normalized semi-classical $|\psi_{SC}(x)|^{2}$ wave function and the quantum $|\psi(x)|^{2}$ wave function. The black dashed line is a fit that serves as a guide for the eye. We observe that $A$ decreases for large velocities, i.e. the higher the velocity the closer the two models are. Below $v=$ 50 m/s, $A$ increases rapidly as $v$ decreases, suggesting a threshold in the validity of the semi-classical model with the parameters considered. (b) Fraction $F$ of atoms exiting the slit as a function of the velocity $v$. Remarkably, almost identical results are obtained for both models.}
    \label{fig:distance}
\end{figure}

In Fig.~\ref{fig:trajectories_semi_classic}, we compare the diffraction patterns obtained by the semi-classical approach and the quantum model. We observe that at high velocity ($v=$ 100 m/s), the two figures are very similar. On the contrary, at low velocity ($v=$ 15 m/s), the semi-classical approach fails to reproduce the numerical result. This observation is highlighted by plotting the following dimensionless quantity 
\begin{equation}
A=\frac{\int_{-\infty}^{\infty} \Big| |\psi_{SC}(x)|^{2} - |\psi(x)|^{2} \Big|\,dx}{\int_{-\infty}^{\infty} |\psi(x)|^{2} \,dx}
\end{equation}
as the function of the velocity $v$ of the incoming atoms. $A$ quantifies the relative difference between the two models. The result is plotted in Fig.~\ref{fig:distance}. Note that for $v \geqslant 50$ m/s, $A$ remains small, meaning that both models give similar results. However, for $v \leqslant 50$ m/s, $A$ increases rapidly as $v$ decreases, indicating a significant discrepancy between the two models, especially in the tails of the distribution. To explain such differences, we can check the validity conditions of the semi-classical approach at the exit of the slit ($t=t_{e}$), which are
\begin{subequations}
\label{eqn:condition}
\begin{equation}
\mathrm{\lambda \frac{\nabla V_{vdW}}{V_{vdW}}} = \frac{3 \lambda}{x \pm a_{s}/2} \ll 1 
\end{equation}
and
\begin{equation}
\frac{1}{2 \pi} \left| \frac{d\lambda}{dx} \right| \ll 1\,.
\end{equation}
\end{subequations}
The first condition means that the de Broglie wavelength $\lambda$ of the particle must be much shorter than the scale on which the potential $V_{vdW}$ changes significantly. The second condition means that the spatial variation of the de Broglie wavelength must be small. For large velocities, both conditions are well met, except close to the surfaces of the slit. For example, at $v$= 100 m/s we have $\lambda (\nabla V_{vdW})/V_{vdW}=0.15 $ and $\frac{1}{2 \pi} | \frac{d\lambda}{dx}| =0.03 $ at 2.7 nm from the surface. This explains why the quantity $A$ saturates at a finite value for large velocities. This feature leads to a tiny difference in the tails of the quantum and semi-classical diffraction patterns. At low velocities, the conditions of Eq.~(\ref{eqn:condition}) are not sufficiently satisfied to find a good agreement between the two models. For instance, at $v=10$ m/s we find  $\lambda (\nabla V_{vdW})/V_{vdW}=0.36$ and $\frac{1}{2\pi} | \frac{d\lambda }{dx}| =0.1 $ at 6.5 nm from the surface. Moreover, this discrepancy at low velocity is certainly increased by the possibility that trajectories other than the classical ones might play a role in the action integral $S(x^{\prime},t_{e})$ \cite{Garashchuk1996}.

In addition to the diffraction pattern, the fraction $F$ of atoms exiting the slit can be studied and compared for the two models. For the semi-classical models, atoms reaching the position $x^{\prime} = \pm a_{s}/2 - r_{min}$ are considered lost, resulting in an effective slit size $\mathrm{W_{eff}}$ (see Fig.~\ref{fig:trajectories_semi_classic}(a)). The fraction $F$ can thus be expressed as $F=\mathrm{W_{eff}/a_{s}}$. In the quantum model, the norm of the wave packet $\psi(x,t)$ is computed to extract the atom losses. The result for the two models is plotted in Fig.~\ref{fig:distance}. Interestingly, we observe that both models give similar results, the difference being at the percent level, with the quantum model giving a slightly higher fraction $F$. We also find that, as expected, the lower the atom velocity, the lower the fraction $F$, meaning that more and more atoms are lost during the propagation inside the slits. This result strengthens our confidence in the validity of the comparison we have made between the two models.

To conclude, Fig.~\ref{fig:distance} shows that the two models give similar results in terms of the fraction of atoms lost. However, in terms of the diffraction pattern (i.e., when atoms are not lost during the propagation), at low velocities and in the tails of the distributions, the semi-classical approach fails and a quantum simulation must be performed.

\section{\label{sec:level5}Conclusion}

In this work, we have presented quantum numerical simulations of atomic diffraction by materials nanogratings. Our simulations are based on a numerically efficient solution of the time-dependent Schrödinger equation. After describing the method and model used, we have demonstrated that our approach goes beyond the semi-classical approach used so far. In particular, we show that our model is able to describe interferometers with low atomic velocities and to capture near-surface effects.

This quantum model could thus be exploited in future experiments with slow atomic or molecular beams. For example, the consequences of long-range quantum reflection \cite{Zhao2016} (i.e., tens of nanometers away from the surface, where no absorption takes place) could be explored. In addition, the increased precision may allow the study of short-range repulsive interactions, which are difficult to describe theoretically with accuracy. This could be done by comparing experiments and numerical simulations based on the present quantum model. Finally, since this model goes beyond the semi-classical approach, it is also relevant to search for possible deviations from Newtonian gravity that could occur at the submicron scale \cite{Klimchitskaya2020}.

\begin{acknowledgments}
We wish to acknowledge Johannes Fiedler for fruitful discussions. This work has been supported by Region Ile-de-France in the framework of DIM SIRTEQ and partially supported by structure fédérative de recherche NAP MOSAIC of the University Sorbonne Paris Nord. N.G. acknowledges funding from the Deutsche Forschungsgemeinschaft (German Research Foundation) under Germany’s Excellence Strategy (EXC-2123 QuantumFrontiers Grants No. 390837967) and through CRC 1227 (DQ-mat) within Projects No. A05, and the German Space Agency at the German Aerospace Center (Deutsche Raumfahrtagentur im Deutschen Zentrum f\"ur Luft- und Raumfahrt, DLR)  with funds provided by the German Federal Ministry of Economic Affairs and Climate Action due to an enactment of the German Bundestag under Grants Nos. 50WM2250A and 50WM2250E (QUANTUS+) and No. 50WM2253A (AI-Quadrat).

\end{acknowledgments}


\revappendix
\section{\label{app:level1}Stationary phase approximation}

\begin{figure}[!h]
    \centering
    \includegraphics[width=\linewidth]{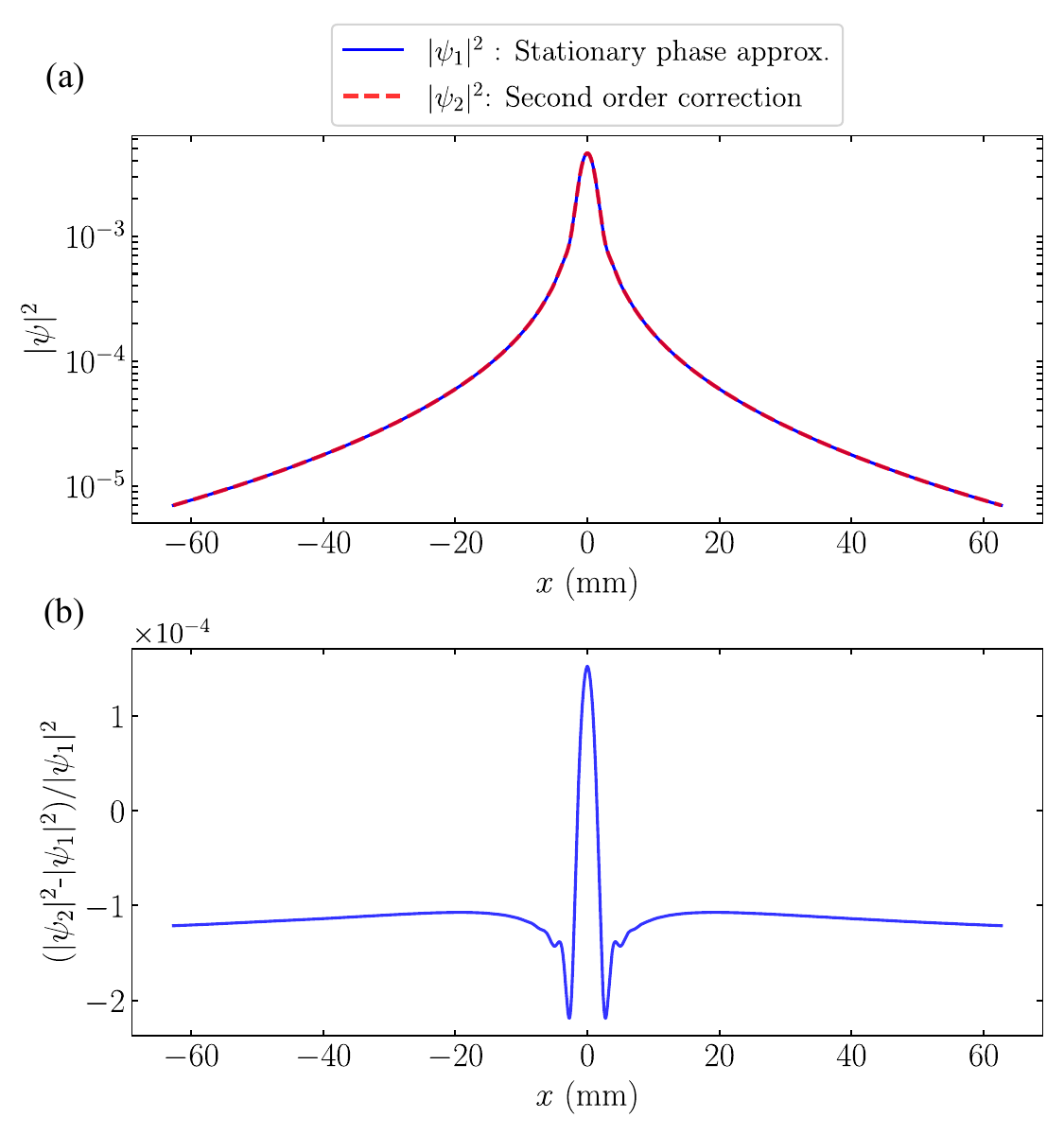}
    \caption{(a) Probability density $|\psi(x,t_f)|^{2}$ in the stationary phase approximation ($|\psi_{1}|^{2}$, blue solid line) and with the second order correction ($|\psi_{2}|^{2}$, red dashed line). For the simulations, the initial velocity is set to $v=$ 15 m/s and the slit geometry is the following: $a_{s}=$ 100 nm and $l_{G}=$ 100 nm. The strength of the Casimir-Polder interaction is fixed at $C_{3}=5.04$ meV.nm$^3$ and the wave function freely expands during $T$ = 21 ms. (b) Relative error between $|\psi_{1}|^{2}$ and $|\psi_{2}|^{2}$. We observe that the maximum relative error is around $2 \times 10^{-4}$, showing that the stationary phase approximation is an excellent approximation here.}
    \label{fig:comparison_stationnary_phase_and_beyond}
\end{figure}

After a propagation time $t_{e}$ inside the slit, the wave function $\psi(x,t_{e})$ evolves freely for a time $T$. Therefore, the wave function at $t_f=t_{e}+T$ is
\begin{equation}
\psi(x,t_f) = \frac{1}{\sqrt{2 \pi}} \int_{-\infty}^{+\infty} e^{-i\frac{\hbar k^2}{2m}T}\, \Tilde{\psi}(k,t_e)\,e^{ikx} \mathrm{d}k\,.
\label{eqn:free_propagation}
\end{equation} 
The global phase is stationary for
\begin{equation}
k=k_{s} = \frac{mx}{ \hbar T}\,.
\end{equation}
With a second order Taylor expansion at $k=k_{s}$, we have
\begin{eqnarray}
\Tilde{\psi}(k,t_e) & \simeq & \Tilde{\psi}(k_{s},t_e) + (k-k_{s})\; \Tilde{\psi}'\!(k_{s},t_e)  \nonumber\\[0.1cm]
& & \qquad\qquad +\; \frac{(k-k_{s})^{2}}{2}\; \Tilde{\psi}''\!(k_{s},t_e)
\end{eqnarray}
and Eq.~(\ref{eqn:free_propagation}) writes  
\begin{align}
\psi(x,t_f) &  \simeq \frac{\Tilde{\psi}(k_{s},t_e)}{\sqrt{2\pi}}  \int_{-\infty}^{+\infty} e^{-i\frac{\hbar k^2}{2m}T}\, e^{ikx} \mathrm{d}k   \nonumber \\
& + \frac{\Tilde{\psi}'\!(k_{s},t_e)}{\sqrt{2 \pi}}  \int_{-\infty}^{+\infty} (k-k_{s}) \,e^{-i\frac{\hbar k^2}{2m}T}\, e^{ikx} \mathrm{d}k   \nonumber \\
& + \frac{\Tilde{\psi}''\!(k_{s},t_e)}{2\sqrt{2 \pi}}  \int_{-\infty}^{+\infty} (k-k_{s})^{2} \,e^{-i\frac{\hbar k^2}{2m}T}\, e^{ikx} \mathrm{d}k
\end{align}
This expression finally yields
\begin{eqnarray}
\psi(x,t_f) & \simeq & \sqrt{\frac{m}{\hbar T}} \;e^{i\left(\frac{m x^2}{2 \hbar T}-\frac{\pi}{4}\right)} \left[\Tilde{\psi}(k_{s},t_e) \right.\nonumber\\
& & \qquad\qquad\left. - i \left(\frac{m}{2\hbar T}\right) \Tilde{\psi}''\!(k_{s},t_e)\right] 
\label{eqn:stationnary_phase_approx_expression}
\end{eqnarray}
The first term in Eq.~(\ref{eqn:stationnary_phase_approx_expression}) corresponds to the so-called stationary phase approximation, and the second term is the first non-zero correction to this approximation. It is by nature a second order term. To quantify the validity of the stationary phase approximation, we plot in Fig.~\ref{fig:comparison_stationnary_phase_and_beyond} the probability density $|\psi(x,t_f)|^{2}$ without and with this second order correction. We observe negligible differences between the two functions, the error bound being below $2 \times 10^{-4}$. In conclusion, the stationary phase approximation yields negligible error and can be used safely in our development.   

\section{\label{app:level2}Estimation of the minimum atom-surface distance}

Throughout the paper we consider Argon (Ar) atoms in the metastable $^{3}P_{2}$ state interacting with a $\mathrm{Si_{3}N_{4}}$ nanograting. For numerical applications and simulations, the minimum atom-surface distance $r_{min}$ must be evaluated. We assume that Ar atoms have a spherical electronic configuration with radius $R$ and that $r_{min} = R$. Using Bohr's model, the energy is given in eV by 
\begin{equation}
E \simeq -13.6 \left( \frac{Z^{*}}{n^{*}} \right)^{2}
\end{equation}
while the radius of the atom is given in atomic units by
\begin{equation}
R \simeq \frac{(n^{*})^{2}}{Z^{*}}
\end{equation}
where $n^{*}$ is the effective quantum number and $Z^{*}$ the effective nuclear charge. For Ar atoms in the $^{3}P_{2}$ state, Slater's rules give $n^{*}$=3.7 and $Z^{*}$=2.05. The numerical application leads to $E=-11.59$ eV for the state $^{3}P_{2}$, close to the measured value $E=-11.55$ eV \cite{Wiese1989}. This confirms the hypothesis of spherical electronic configuration for Ar. For the radius we find $R=r_{min}$ = 0.35 nm, which is the value used throughout the paper.



\begin{thebibliography}{99}
\bibitem{Geiger2020} R. Geiger, A. Landragin, S. Merlet, F. Pereira Dos Santos, High-accuracy inertial measurements with cold-atom sensors, \href{https://pubs.aip.org/avs/aqs/article-abstract/2/2/024702/997275/High-accuracy-inertial-measurements-with-cold-atom?redirectedFrom=fulltext}{AVS Quantum Sci. \textbf{2}, 024702 (2020).}

\bibitem{Machluf2013} S. Machluf, Y. Japha and R. Folman, Coherent Stern–Gerlach momentum splitting on an atom chip, \href{https://www.nature.com/articles/ncomms3424}{Nat Commun \textbf{4}, 2424 (2013).}  

\bibitem{Cronin2009} A. D. Cronin, J. Schmiedmayer, and D. E. Pritchard, Optics and interferometry with atoms and molecules, \href{https://journals.aps.org/rmp/abstract/10.1103/RevModPhys.81.1051}{Rev. Mod. Phys. \textbf{81}, 1051 (2009).} 

\bibitem{Tino2021} G. M Tino, Testing gravity with cold atom interferometry: results and prospects, \href{https://iopscience.iop.org/article/10.1088/2058-9565/abd83e}{Quantum Sci. Technol. \textbf{6}, 024014 (2021).}

\bibitem{Dutta2016} I. Dutta, D. Savoie, B. Fang, B. Venon, C.L. Garrido Alzar, R. Geiger, and A. Landragin, Continuous Cold-Atom Inertial Sensor with 1 nrad/sec Rotation Stability, \href{https://journals.aps.org/prl/abstract/10.1103/PhysRevLett.116.183003}{Phys. Rev. Lett. \textbf{116}, 183003 (2016).} 

\bibitem{Buhmann2012} S. Y. Buhmann, \textit{Dispersion Forces I: Macroscopic quantum electrodynamics and ground-state Casimir, Casimir– Polder and van der Waals forces} (Springer, Heidelberg, 2012).

\bibitem{Bruch1997} L. W. Bruch, M. W. Cole, and E. Zaremba, \textit{Physical Adsorption: Forces and Phenomena} (Clarendon, Oxford, 1997).

\bibitem{Fiedler2022} J. Fiedler, B. Holst, An atom passing through a hole in a dielectric membrane: impact of dispersion forces on mask-based matter-wave lithography, \href{https://iopscience.iop.org/article/10.1088/1361-6455/ac4b41/meta}{J. Phys. B: At. Mol. Opt. Phys. \textbf{55} 025401 (2022).}  

\bibitem{Perreault2005} J. D. Perreault, A. D. Cronin, and T. A. Savas, Using atomic diffraction of Na from material gratings to measure atom-surface interactions, \href{https://journals.aps.org/pra/abstract/10.1103/PhysRevA.71.053612}{Phys. Rev. A \textbf{71}, 053612 (2005).} 

\bibitem{Lepoutre2009} S. Lepoutre, H. Jelassi, V. P. A. Lonij, G. Trénec, M. Büchne1, A. D. Cronin and J. Vigué, Dispersive atom interferometry phase shifts due to atom-surface interactions, \href{https://iopscience.iop.org/article/10.1209/0295-5075/88/20002}{EPL, \textbf{88}, 20002 (2009).} 

\bibitem{Grisenti1999} R. E. Grisenti, W. Schöllkopf, J. P. Toennies, G. C. Hegerfeldt, and T. Köhler, Determination of Atom-Surface van der Waals Potentials from Transmission-Grating Diffraction Intensities, \href{https://journals.aps.org/prl/abstract/10.1103/PhysRevLett.83.1755}{Phys. Rev. Lett. \textbf{83}, 1755 (1999).} 

\bibitem{Bruhl2002} R. Brühl, P. Fouquet, R. E. Grisenti, J. P. Toennies, G. C. Hegerfeldt, T. Köhler, M. Stoll and C. Walter, The van der Waals potential between metastable atoms and solid surfaces: Novel diffraction experiments vs. theory, \href{https://iopscience.iop.org/article/10.1209/epl/i2002-00202-4}{EPL 59 357 (2002).} 

\bibitem{Brand2015} C. Brand, M. Sclafani, C. Knobloch, Y. Lilach, T. Juffmann, J. Kotakoski, C. Mangler, A. Winter, A. Turchanin, J. Meyer, O. Cheshnovski, M. Arndt, An atomically thin matter-wave beamsplitter, \href{https://www.nature.com/articles/nnano.2015.179}{Nat. Nanotechnol  \textbf{10}, 845-848 (2015).} 

\bibitem{Garcion2021} C. Garcion, N. Fabre, H. Bricha, F. Perales, S. Scheel, M. Ducloy, and G. Dutier, Intermediate-Range Casimir-Polder Interaction Probed by High-Order Slow Atom Diffraction, \href{https://journals.aps.org/prl/abstract/10.1103/PhysRevLett.127.170402}{Phys. Rev. Lett. \textbf{127}, 170402 (2021).} 

\bibitem{Gregoire2015} M. D. Gregoire, I. Hromada, W. F. Holmgren, R. Trubko, A. D. Cronin, Measurements of the ground-state polarizabilities of Cs, Rb, and K using atom interferometry, \href{https://journals.aps.org/pra/abstract/10.1103/PhysRevA.92.052513}{Phys. Rev. A \textbf{92}, 052513 (2015).} 

\bibitem{Hackermuller2007} L. Hackermuller, K. Hornberger, S. Gerlich, M. Gring, H. Ulbricht, and M. Arndt, Optical polarizabilities of large molecules measured in near-field interferometry, \href{https://link.springer.com/article/10.1007/s00340-007-2873-6}{Appl. Phys. B \textbf{89}, 469 (2007).} 

\bibitem{Trubko2015} R. Trubko, J. Greenberg, M. T. St. Germaine, M. D. Gregoire, W. F. Holmgren, I. Hromada and A. D. Cronin, Atom Interferometer Gyroscope with Spin-Dependent Phase Shifts Induced by Light near a Tune-Out Wavelength, \href{https://journals.aps.org/prl/abstract/10.1103/PhysRevLett.114.140404}{Phys. Rev. Lett. \textbf{114}, 140404 (2015).} 

\bibitem{Cronin2004} A. D. Cronin and J. D. Perreault, Phasor analysis of atom diffraction from a rotated material grating, \href{https://journals.aps.org/pra/abstract/10.1103/PhysRevA.70.043607}{Phys. Rev. A \textbf{70}, 043607 (2004).} 

\bibitem{Nimmrichter2008} S. Nimmrichter and K. Hornberger, Theory of near-field matter-wave interference beyond the eikonal approximation, \href{https://journals.aps.org/pra/abstract/10.1103/PhysRevA.78.023612}{Phys. Rev. A \textbf{78}, 023612 (2008).} 

\bibitem{Herwerth2013} B. Herwerth, M. DeKieviet, J. Madroñero and S. Wimberger, Quantum reflection from an oscillating surface, \href{https://iopscience.iop.org/article/10.1088/0953-4075/46/14/141002}{J. Phys. B: At. Mol. Opt. Phys. \textbf{46} 141002 (2013).} 

\bibitem{Galiffi2017} E. Galiffi, C. Sünderhauf, M. DeKieviet and S. Wimberger, Two-dimensional simulation of quantum reflection,  \href{https://iopscience.iop.org/article/10.1088/1361-6455/aa66e2/meta}{J. Phys. B: At. Mol. Opt. Phys. \textbf{50} 095001 (2017).}

\bibitem{Mandel1995} L. Mandel and E. Wolf, \textit{Optical Coherence and Quantum Optics} (Cambridge University Press, 1995).

\bibitem{Sipe1984} J. M. Wylie and J. E. Sipe, Quantum electrodynamics near an interface, \href{https://journals.aps.org/pra/abstract/10.1103/PhysRevA.30.1185}{Phys. Rev. A \textbf{30}, 1185 (1984).}

\bibitem{Sipe1985} J. M. Wylie and J. E. Sipe, Quantum electrodynamics near an interface. II, \href{https://journals.aps.org/pra/abstract/10.1103/PhysRevA.32.2030}{Phys. Rev. A \textbf{32}, 2030 (1985).}

\bibitem{C3_calcul} The value of the $C_{3}$ coefficient is calculated in the framework of Lifshitz theory, assuming a semi-infinite plan and inferring the spectral response of the $\mathrm{Si_{3}N_{4}}$ material through the two following references: \href{https://iopscience.iop.org/article/10.1149/1.2403440}{H. R. Philipp, Optical Properties of Silicon Nitride,
J. Electrochem. Soc. \textbf{120}, 295 (1973)} and K. Luke, Y. Okawachi, M. R. E. Lamont, A. L. Gaeta, and M. Lipson, \href{https://opg.optica.org/ol/fulltext.cfm?uri=ol-40-21-4823&id=331311}{Broadband mid-infrared frequency comb generation in a $\mathrm{Si_{3}N_{4}}$ microresonator, Opt. Lett. \textbf{40}, 4823 (2015). }

\bibitem{Bryk1999} P. Bryk, S. Sokolowski, D. Henderson, Some aspects of the adsorption of a Lennard-Jones gas on a rough surface, \href{https://pubs.aip.org/aip/jcp/article-abstract/110/1/15/474914/Some-aspects-of-the-adsorption-of-a-Lennard-Jones}{J. Chem. Phys. \textbf{110}, 15–17 (1999).}

\bibitem{Maury2016} A. Maury, M. Donaire, M.P. Gorza, A. Lambrecht, and R. Guérout, Surface-modified Wannier-Stark states in a one-dimensional optical lattice, \href{https://journals.aps.org/pra/abstract/10.1103/PhysRevA.94.053602}{Phys. Rev. A \textbf{94}, 053602 (2016).}

\bibitem{Lonij2009} V. P. A. Lonij, W. F. Holmgren, A. D. Cronin, Magic ratio of window width to grating period for van der Waals potential measurements using material gratings, \href{https://journals.aps.org/pra/abstract/10.1103/PhysRevA.80.062904}{Phys. Rev. A \textbf{80}, 062904 (2009).}

\bibitem{Vassen2012} W. Vassen, C. Cohen-Tannoudji, M. Leduc, D. Boiron, C. I. Westbrook, A. Truscott, K. Baldwin, G. Birkl, P. Cancio, and M. Trippenbach, Cold and trapped metastable noble gases,  \href{https://journals.aps.org/rmp/abstract/10.1103/RevModPhys.84.175}{Rev. Mod. Phys. \textbf{84}, 175 (2012).} 

\bibitem{Kosloff1986} R. Kosloff and D. Kosloff, Absorbing boundaries for wave propagation problems, \href{https://www.sciencedirect.com/science/article/pii/0021999186901993}{J. Comput. Phys. \textbf{63}, 363-376 (1986).}

\bibitem{Hussain2000} A. N. Hussain and G. Roberts, Procedure for absorbing time-dependent wave functions at low kinetic energies and large bandwidths, \href{https://journals.aps.org/pra/abstract/10.1103/PhysRevA.63.012703}{Phys. Rev. A \textbf{63}, 012703 (2000).}

\bibitem{Feit1982} M. D. Feit, J. A. Fleck, and A. Steiger, Solution of the Schrödinger equation by a spectral method, \href{https://www.sciencedirect.com/science/article/pii/0021999182900912}{J. Comput. Phys. \textbf{47}, 412-433 (1982).}

\bibitem{Born1999} M. Born and E. Wolf, \textit{Principles of Optics: Electromagnetic Theory of Propagation, Interference and Diffraction of Light} (Cambridge University Press, 7th edition, 1999).

\bibitem{Henkel1994} C. Henkel, J.-Y. Courtois, A. Aspect, Atomic diffraction by a thin phase grating, \href{https://arxiv.org/abs/quant-ph/0408154}{J. Phys. II France \textbf{4}, 1955-74 (1994).}

\bibitem{Garashchuk1996} S Garashchuk and D. Tannor, Calculation of autocorrelation functions using the Wigner representation of quantum mechanics, \href{https://www.sciencedirect.com/science/article/pii/S0009261496011839}{Chem. Phys. Lett. \textbf{263}, 324-330 (1996).}

\bibitem{Zhao2016} B.S Zhao, Z. Zhang, and W. Schöllkopf, Universal diffraction of atoms and molecules from a quantum reflection grating, \href{https://www.science.org/doi/full/10.1126/sciadv.1500901}{Sci. Adv. \textbf{2}, e1500901 (2016).}

\bibitem{Klimchitskaya2020} G.L. Klimchitskaya, P. Kuusk and V.M. Mostepanenko, Constraints on non-Newtonian gravity and axionlike particles from measuring the Casimir force in nanometer separation range, \href{https://journals.aps.org/prd/abstract/10.1103/PhysRevD.101.056013}{Phys. Rev. D \textbf{101}, 056013 (2020).}  

\bibitem{Wiese1989} W. L. Wiese, J. W. Brault, K. Danzmann, V. Helbig, and M. Kock, Unified set of atomic transition probabilities for neutral argon, \href{https://journals.aps.org/pra/abstract/10.1103/PhysRevA.39.2461}{Phys. Rev. A \textbf{39}, 2461 (1989).}


\end{thebibliography}
\end{document}